\documentclass[11pt]{article}%
\usepackage{amsfonts}
\usepackage{amsmath}
\usepackage{amssymb}
\usepackage{graphicx}%
\providecommand{\U}[1]{\protect\rule{.1in}{.1in}}

\newenvironment{proof}[1][Proof]{\noindent\textbf{#1.} }{\ \rule{0.5em}{0.5em}}
\begin{document}

\title{The Reconstruction Problem and Weak Quantum Values }
\author{Maurice A. de Gosson\\University of Vienna\\Faculty of Mathematics, NuHAG\\A-1090 Vienna
\and Serge M. de Gosson\\Swedish Social Insurance Agency\\Department for Analysis and Forecasts\\S-103 51 Stockholm}
\maketitle

\begin{abstract}
Quantum Mechanical weak values are an interference effect measured by the
cross-Wigner transform $\mathcal{W}(\phi,\psi)$ of the post-and preselected
states, leading to a complex quasi-distribution $\rho_{\phi,\psi}(x,p)$ on
phase space. We show that the knowledge of $\rho_{\phi,\psi}(z)$ and of one of
the two functions $\phi,\psi$ unambiguously determines the other, thus
generalizing a recent reconstruction result of Lundeen and his collaborators.

\end{abstract}

\section{Introduction}

In 1958 W. Pauli \cite{pauli} mentions the problem of the reconstruction of a
quantum state knowing its position and momentum; this conjecture was later
disproved; see H. Reichenbach's book \cite{Reichenbach}; also Corbett
\cite{corbett} for a review of the Pauli problem. However, in \cite{lund}
Lundeen and his coworkers consider the following experiment: a weak
measurement of position is performed on a particle; thereafter a strong
measurement of momentum is made. This allows them to effectively reconstruct
the wavefunction $\psi(x)$ pointwise by scanning through all values of $x$.
The aim of the present paper is to show that, more generally, the wavefunction
can be reconstructed from the knowledge of a complex probability distribution
$\rho_{\phi,\psi}(x,p)$ on phase space, related to the notion of weak values,
and expressed in the terms of the cross-Wigner transform $\mathcal{W}%
(\phi,\psi)$ of the pre- and postselected states. Our proof is based on the
fact that the knowledge of $\mathcal{W}(\phi,\psi)$ and of one of the two
functions $\phi,\psi$ allows to determine uniquely and unambiguously (in
infinitely many ways) the other function. This is of course in strong contrast
with the case of the usual Wigner distribution $\mathcal{W}\psi=\mathcal{W}%
(\psi,\psi)$ whose knowledge only determines $\psi$ up to a factor with
modulus one.

In what follows the position and momentum vectors are $x=(x_{1},...,x_{n})$
and $p=(p_{1},...,p_{n})$, respectively; $px=p_{1}x_{1}+\cdot\cdot\cdot
+p_{n}x_{n}$ is their scalar product. We write $\mathrm{d}^{n}x=\mathrm{d}%
x_{1}\mathrm{d}x_{2}\cdot\cdot\cdot\mathrm{d}x_{n}$, and all the integrations
are performed over $n$-dimensional space $\mathbb{R}^{n}$. The $\hbar$-Fourier
transform of a function $\psi$ is defined by
\[
F\psi(p)=\left(  \frac{1}{2\pi\hbar}\right)  ^{n}\int e^{-\frac{\mathrm{i}%
}{\hbar}px}\psi(x)\mathrm{d}^{n}x.
\]

\section{Weak Values and the Cross-Wigner Transform}

Let $\widehat{A}$ be a quantum observable associated to a function $A$ by the
Weyl correspondence: $\widehat{A}\overset{\mathrm{Weyl}}{\longleftrightarrow
}A$ \cite{Birk,Birkbis,Littlejohn} and $\phi,\psi$ two non-orthogonal states.
Aharonov and coworkers \cite{abl,av,avphysa,apt,ahro} have introduced and
studied the notion of weak measurement and the weak value of an observable
(also see \cite{hiley}). The \emph{weak value} of $\widehat{A}$ with respect
to the pair $(\phi,\psi)$ is the complex number%
\begin{equation}
\langle\widehat{A}\rangle_{\mathrm{weak}}^{\phi,\psi}=\frac{\langle
\phi|\widehat{A}|\psi\rangle}{\langle\phi|\psi\rangle}\text{ \ , \ }%
\langle\phi|\psi\rangle\neq0.\label{abl}%
\end{equation}
According to%
\begin{equation}
\langle\widehat{A}\rangle_{\mathrm{weak}}^{\phi,\psi}=\int A(x,p)\rho
_{\phi,\psi}(x,p)d^{n}xd^{n}p\label{rho}%
\end{equation}
When $\phi=\psi$ this is the usual average value $\langle\widehat{A}%
\rangle^{\psi}$ of $\widehat{A}$ in the state $\psi$. The physical
interpretation of the weak value is the following (see Jozsa \cite{jozsa} and
Parks and Gray \cite{pg} for concise and clear expositions): viewing
$|\psi\rangle$ as a preselected state and $\langle\phi|$ as a postselected
state, if we couple a measuring device whose pointer has position coordinate
$x$ to the system and subsequently measure that coordinate then the mean value
of the pointer is
\[
\left\langle \widehat{x}\right\rangle =g\operatorname{Re}\langle
\widehat{A}\rangle_{\mathrm{weak}}^{\phi,\psi}%
\]
if the coupling interaction is the standard von Neumann interaction
Hamiltonian $\widehat{H}=g\widehat{A}p$. In addition the mean of the pointer
momentum is given by
\[
\left\langle \widehat{p}\right\rangle =2\frac{g}{\hbar}v\operatorname{Im}%
\langle\widehat{A}\rangle_{\mathrm{weak}}^{\phi,\psi}.
\]
(This needs sufficiently weak interaction, see the discussions in Duck et al.
\cite{duck} and in Parks et al. \cite{pcs}).

In a recent work \cite{gogo} we have shown that the weak value can be
calculated by averaging $A$ over the complex phase space function%
\begin{equation}
\rho_{\phi,\psi}(x,p)=\frac{\mathcal{W}(\phi,\psi)(x,p)}{\langle\phi
|\psi\rangle} \label{rofipsi}%
\end{equation}
where
\begin{equation}
\mathcal{W}(\phi,\psi)(x,p)=\left(  \tfrac{1}{2\pi\hbar}\right)  ^{n}\int
e^{-\frac{\mathrm{i}}{\hbar}py}\phi^{\ast}(x+\tfrac{1}{2}y)\psi(x-\tfrac{1}%
{2}y)\mathrm{d}^{n}y \label{wigner}%
\end{equation}
is the cross-Wigner transform (we are conjugating $\phi$ and not $\psi$ in
order to be consistent with the bra-ket notation; in most mathematical texts
the transform defined by (\ref{wigner}) would be denoted by $\mathcal{W}%
(\psi,\phi)$). The function $\rho_{\phi,\psi}$ satisfies the marginal
conditions%
\begin{equation}
\int\rho_{\phi,\psi}(x,p)\mathrm{d}^{n}p=\frac{\phi^{\ast}(x)\psi(x)}%
{\langle\phi|\psi\rangle}\text{ , }\int\rho_{\phi,\psi}(x,p)\mathrm{d}%
^{n}x=\frac{[F\phi(p)]^{\ast}F\psi(p)}{\langle\phi|\psi\rangle};
\label{marginals}%
\end{equation}
and hence in particular%
\[
\int\rho_{\phi,\psi}(x,p)\mathrm{d}^{n}x\mathrm{d}^{n}p=1.
\]
The function $\rho_{\phi,\psi}$ can thus be viewed as a complex
quasi-probability density on phase space; its real and imaginary parts
moreover satisfy%
\[
\int\operatorname{Re}\rho_{\phi,\psi}(x,p)\mathrm{d}^{n}x\mathrm{d}%
^{n}p=1\text{ \ , \ }\int\operatorname{Im}\rho_{\phi,\psi}(x,p)\mathrm{d}%
^{n}x\mathrm{d}^{n}p=1\text{ }%
\]

The appearance of the cross-Wigner function is characteristic of interference
phenomena, and suggests the following interpretation of weak
values\footnote{But this interpretation is far from being unanimously shared
in the scientific community, and has led to an ongoing epistemological debate
on the back-action of future on the past.} \cite{av,avphysa,apt}. Assume that
we measure an observable $\widehat{A}$ an initial time $t_{\mathrm{in}}$ and
that a non-degenerate eigenvalue was found: $|\psi(t_{\mathrm{in}}%
)\rangle=|\widehat{A}=a\rangle$ (the pre-selected state); similarly at a final
time $t_{\mathrm{fin}}$ a measurement of another observable $\widehat{B}$
yields $|\phi(t_{\mathrm{fin}})\rangle=|\widehat{B}=b\rangle$ (the
post-selected state). Choose now some intermediate time $t$ : $t_{\mathrm{in}%
}<t<t_{\mathrm{fin}}$. Following the time-symmetric approach to quantum
mechanics, at this intermediate time the system is described by the \emph{two}
wavefunctions%
\begin{equation}
\psi=\psi_{t}=U_{t,t_{\mathrm{in}}}^{H_{\mathrm{in}}}\psi(t_{\mathrm{in}%
})\text{ \ , \ }\phi=\phi_{t}=U_{t,t_{\mathrm{fin}}}^{H_{\mathrm{fin}}}%
\phi(t_{\mathrm{fin}})\label{t12}%
\end{equation}
where $U_{t,t^{\prime}}^{H}=e^{-\mathrm{i}\widehat{H}(t-t^{\prime}%
)/\hbar\text{ }}$ is the Schr\"{o}dinger unitary\ evolution operator if
$\widehat{H}$ is the quantum Hamiltonian). Notice that $\phi_{t}$ travels
\emph{backwards} in time since $t<t_{\mathrm{fin}}$. The situation is thus the
following: at any time $t^{\prime}<t$ the system under consideration is in the
state $|\psi_{t^{\prime}}\rangle=U_{t^{\prime},t_{\mathrm{in}}}^{H}%
|\psi(t_{\mathrm{in}})\rangle$ and has Wigner distribution $\mathcal{W}%
\psi_{t^{\prime}}$; at any time $t^{\prime\prime}>t$ the system is in the
state $|\phi_{t^{\prime\prime}}\rangle=U_{t^{\prime\prime},t_{\mathrm{fin}}%
}^{H}|\phi(t_{\mathrm{fin}})\rangle$ and has Wigner distribution
$\mathcal{W}\phi_{t^{\prime}}$. But at time $t$ it is the superposition
$|\psi_{t}\rangle+|\phi_{t}\rangle$ of both states, and the Wigner
distribution
\[
\mathcal{W}(\phi_{t}+\psi_{t})=\mathcal{W}(\phi_{t}+\psi_{t},\phi_{t}+\psi
_{t})
\]
of this state is
\begin{equation}
\mathcal{W}(\phi_{t}+\psi_{t})=\mathcal{W}\phi_{t}+\mathcal{W}\psi
_{t}+2\operatorname{Re}\mathcal{W}(\phi_{t},\psi_{t}).\label{wpsifi}%
\end{equation}
This equality shows the emergence at time $t$ of the interference term
$2\operatorname{Re}\mathcal{W}(\phi_{t},\psi_{t})=2\operatorname{Re}%
\mathcal{W}(\phi,\psi)$, signalling a strong interaction between the states
$|\psi_{t}\rangle=|\psi\rangle$ and $|\phi_{t}\rangle=|\phi\rangle$.

\section{The Reconstruction Problem}

\subsection{An example}

As briefly explained in the Introduction Lundeen and his coworkers \cite{lund}
consider the following experiment: a weak measurement of position is performed
on a particle; this amounts to applying the projection operator $\widehat{\Pi
}_{x}=|x\rangle\langle x|$ to the pre-selected state $|\psi\rangle$, which
yields $\widehat{\Pi}_{x}|\psi\rangle=\psi(x)|x\rangle$. Thereafter a strong
measurement of momentum is made, yielding a value $p_{0}$; the result of the
weak measurement is thus%
\begin{equation}
\langle\widehat{\Pi}_{x}\rangle_{\mathrm{weak}}^{\phi,\psi}=\frac
{e^{\frac{\mathrm{i}}{\hbar}p_{0}x}\psi(x)}{F\psi(p_{0})}. \label{pix1}%
\end{equation}
This allows the reconstruction of the function $\psi$ from $\langle
\widehat{\Pi}_{x}\rangle_{\mathrm{weak}}^{\phi,\psi}$:%
\begin{equation}
\psi(x)=ke^{-\frac{\mathrm{i}}{\hbar}p_{0}x}\langle\widehat{\Pi}_{x}%
\rangle_{\mathrm{weak}}^{\phi,\psi}\text{ \ , \ }k=F\psi(p_{0}).
\label{lundeen}%
\end{equation}

Let us retrieve this result using the Wigner formalism developed above.
Obviously, the operator $\widehat{A}$ is here the projector $\widehat{\Pi}%
_{x}$ whose analytical expression is given by%
\begin{equation}
\widehat{\Pi}_{x}\psi(y)=\psi(x)\delta(x-y).\label{proj}%
\end{equation}
Choosing for the post-selected state $\phi$ the normalized momentum
wavefunction
\[
\phi_{p_{0}}(x)=\left(  2\pi\hbar\right)  ^{-n/2}e^{\frac{\mathrm{i}}{\hbar
}p_{0}x}%
\]
we have
\begin{align*}
\mathcal{W}(\psi,\phi_{p_{0}})(x,p) &  =\left(  \tfrac{1}{2\pi\hbar}\right)
^{3n/2}\int e^{-\frac{\mathrm{i}}{\hbar}py}\psi^{\ast}(x+\tfrac{1}%
{2}y)e^{\frac{\mathrm{i}}{\hbar}p_{0}(x-\frac{1}{2}y)}\mathrm{d}^{n}y\\
&  =\left(  \tfrac{1}{2\pi\hbar}\right)  ^{3n/2}e^{\frac{\mathrm{i}}{\hbar
}p_{0}x}\int e^{-\frac{\mathrm{i}}{\hbar}(p+\frac{1}{2}p_{0})y}\psi^{\ast
}(x+\tfrac{1}{2}y)\mathrm{d}^{n}y
\end{align*}
which --after the change of variables $x^{\prime}=x+\frac{1}{2}y$ and
integrating-- becomes%
\[
\mathcal{W}(\psi,\phi_{p_{0}})(x,p)=\left(  \tfrac{1}{\pi\hbar}\right)
^{n}e^{\frac{2\mathrm{i}}{\hbar}(p-p_{0})x}F\psi(2p-p_{0}).
\]
Taking into account the fact that $\langle\phi_{p_{0}}|\psi\rangle=F\psi
(p_{0})$ this yields
\begin{equation}
\rho_{\psi,p_{0}}(x,p)=\left(  \frac{1}{\pi\hbar}\right)  ^{n}e^{\frac
{2\mathrm{i}}{\hbar}(p-p_{0})x}\frac{F\psi(2p-p_{0})}{F\psi(p_{0}%
)}.\label{roo}%
\end{equation}
In view of Eqn. (\ref{proj}) the classical observable $\Pi_{x}%
\overset{\mathrm{Weyl}}{\longleftrightarrow}\widehat{\Pi}_{x}$ is given by
$\Pi_{x}(x^{\prime},p^{\prime})=\delta(x^{\prime}-x)$ and hence%
\begin{align*}
\langle\widehat{\Pi}_{x}\rangle_{\mathrm{weak}}^{\phi_{p_{0}},\psi} &
=\int\rho_{\psi,p_{0}}(x^{\prime},p^{\prime})\delta(x^{\prime}-x)\mathrm{d}%
^{n}p^{\prime}\mathrm{d}^{n}x^{\prime}\\
&  =\int\rho_{\psi,p_{0}}(x,p^{\prime})\mathrm{d}^{n}p^{\prime}\\
&  =\frac{\phi_{p_{0}}^{\ast}(x)\psi(x)}{\langle\phi_{p_{0}}|\psi\rangle}%
\end{align*}
where the last equality is a consequence of the first marginal distribution
property (\ref{marginals}); this is precisely the expression (\ref{pix1}) of
the weak value of $\widehat{\Pi}_{x}$.

We note that the difficulties inherent to the theory of measurement of
continuous conjugate variables (such as position and momentum) are real, and
far from being resolved. For a mathematically rigorous approach, using the
properties of the metaplectic group, see Weigert and Wilkinson \cite{wewi}.

\subsection{A general reconstruction formula}

Formula (\ref{pix1}) shows that we can reconstruct the whole wavefunction
$\psi$ by scanning the weak measurements of the projection operator
$\widehat{\Pi}_{x}$ through $x$. We are going to show that, more generally, a
quantum state $\psi$ can always be reconstructed from the knowledge of the
complex quasi-distribution $\rho_{\phi_{p},\psi}$, or, equivalently from the
knowledge of the cross-interference term $\mathcal{W}(\phi,\psi)$. Let us
begin by introducing some notation. Let $z=(x,p)$ and $z_{0}=(x_{0},p_{0})$ be
an arbitrary phase space point ,and define the Grossmann--Royer operator
\cite{gros,roy} $\widehat{T}_{\text{GR}}(z_{0})$ by
\begin{equation}
\widehat{T}_{\text{GR}}(z_{0})\psi(x)=e^{\frac{2\mathrm{i}}{\hbar}%
p_{0}(x-x_{0})}\psi(2x_{0}-x).\label{groroy1}%
\end{equation}
It is, up to the complex exponential factor in front of $\psi(2x_{0}-x)$ a
reflection operator, in fact $\widehat{T}_{\text{GR}}(z_{0})\widehat{T}%
_{\text{GR}}(z_{0})=\widehat{I}_{\mathrm{d}}$ (the identity operator). It is
in addition a unitary self-adjoint operator: $\widehat{T}_{\text{GR}}%
(z_{0})^{\ast}=\widehat{T}_{\text{GR}}(z_{0})^{-1}=\widehat{T}_{\text{GR}%
}(z_{0})$. A remarkable fact is that the cross-Wigner transform is related to
$\widehat{T}_{\text{GR}}(z)$ by the simple formula%
\begin{equation}
\mathcal{W}(\phi,\psi)(z)=\left(  \tfrac{1}{\pi\hbar}\right)  ^{n}%
\langle\widehat{T}_{\text{GR}}(z)\phi|\psi\rangle\label{wigroyer}%
\end{equation}
(see \cite{Birkbis}, Chapter 9). Using the following well-known formula
(\textquotedblleft Moyal identity\textquotedblright, \cite{Birkbis}, Chapter
9):%
\begin{equation}
\int\mathcal{W}(\phi,\psi)^{\ast}(z)\mathcal{W}(\phi^{\prime},\psi^{\prime
})(z)\mathrm{d}^{2n}z=\left(  \tfrac{1}{2\pi\hbar}\right)  ^{n}\langle
\phi|\phi^{\prime}\rangle\langle\psi|\psi^{\prime}\rangle\label{wigwig}%
\end{equation}
and the Grossmann--Royer formalism we prove that the knowledge of the
cross-Wigner transform $\mathcal{W}(\phi,\psi)$, and of that one of the two
functions $\phi,\psi$ uniquely determines the other. Moreover, this function
can be written in terms of an arbitrary square-integrable auxiliary function
$\gamma$. 

\noindent\textbf{Proposition}. Let $(\phi,\gamma)$ be a pair of square
integrable functions such that $\langle\gamma|\phi\rangle\neq0$. We have%
\begin{align}
\phi(x)  &  =\frac{2^{n}}{\langle\psi|\gamma\rangle}\int\mathcal{W}(\phi
,\psi)(z_{0})\widehat{T}_{\text{GR}}(z_{0})\gamma(x)\mathrm{d}^{2n}%
z_{0}\label{bf1}\\
\psi(x)  &  =\frac{2^{n}}{\langle\phi|\gamma\rangle}\int\mathcal{W}(\phi
,\psi)^{\ast}(z_{0})\widehat{T}_{\text{GR}}(z_{0})\gamma(x)\mathrm{d}%
^{2n}z_{0}. \label{bf2}%
\end{align}

\begin{proof}
We begin with a preliminary remark: both integrals in the formulas above are
absolutely convergent. In fact, taking $\phi=\phi^{\prime}$ and $\psi
=\psi^{\prime}$ in Moyal's identity (\ref{wigwig}) we see that $\mathcal{W}%
(\phi,\psi)$ is square integrable. In view of the Cauchy--Schwarz inequality
we have%
\[
\int|\mathcal{W}(\phi,\psi)(z_{0})||\widehat{T}_{\text{GR}}(z_{0}%
)\gamma(x)|\mathrm{d}^{2n}z_{0}\leq||\mathcal{W}(\phi,\psi)||_{L^{2}}%
||\gamma||_{L^{2}}<+\infty
\]
for the integrals in (\ref{bf1}), (\ref{bf2}). We next observe that both
formulas (\ref{bf1}) and (\ref{bf2}) are equivalent and obtained from each
other by swapping $\phi$ and $\psi$ and noting that $\mathcal{W}(\psi
,\phi)^{\ast}=\mathcal{W}(\phi,\psi)$. Let us prove (\ref{bf1}). Let us denote
by $\chi(x)$ the right hand side of (\ref{bf1}):
\[
\chi(x)=\frac{2^{n}}{\langle\psi|\gamma\rangle}\int\mathcal{W}(\phi
,\psi)(z_{0})\widehat{T}_{\text{GR}}(z_{0})\gamma(x)\mathrm{d}^{2n}z_{0}.
\]
We are going to show that $\langle\chi|\theta\rangle=\langle\phi|\theta
\rangle$ for every element $\theta$ of the Schwartz space $\mathcal{S}%
(\mathbb{R}^{n})$; it will follow that we have $\chi=\phi$ almost everywhere,
which proves formula (\ref{bf1}). We have%
\[
\langle\chi|\theta\rangle=\frac{2^{n}}{\langle\psi|\gamma\rangle}%
\int\mathcal{W}(\phi,\psi)(z)\langle\widehat{T}_{\text{GR}}(z)\gamma
|\theta\rangle\mathrm{d}^{2n}z.
\]
In view of formula (\ref{wigroyer}) we have%
\[
\langle\widehat{T}_{\text{GR}}(z)\gamma|\theta\rangle=(\pi\hbar)^{n}%
\mathcal{W}(\gamma,\theta)(z)
\]
and hence%
\begin{align*}
\langle\chi|\theta\rangle &  =\frac{(2\pi\hbar)^{n}}{\langle\psi|\gamma
\rangle}\int\mathcal{W}(\phi,\psi)(z)\mathcal{W}(\gamma,\theta)(z)\mathrm{d}%
^{2n}z\\
&  =\frac{(2\pi\hbar)^{n}}{\langle\psi|\gamma\rangle}\int\mathcal{W}(\psi
,\phi)^{\ast}(z)\mathcal{W}(\gamma,\theta)(z)\mathrm{d}^{2n}z.
\end{align*}
Applying Moyal's identity (\ref{wigwig}) to the last integral we get%
\[
\int\mathcal{W}(\psi,\phi)^{\ast}(z)\mathcal{W}(\gamma,\theta)\mathrm{d}%
^{2n}z=\left(  \tfrac{1}{2\pi\hbar}\right)  ^{n}\langle\psi|\gamma
\rangle\langle\phi|\theta\rangle
\]
and hence $\langle\chi|\theta\rangle=\langle\phi|\theta\rangle$. 
\end{proof}

Since the cross-Wigner transform $\mathcal{W}(\phi,\psi)$ and the weak value
$\rho_{\phi,\psi}(x,p)$ are equal up to a factor $\langle\phi|\psi\rangle$
(formula (\ref{rofipsi})), it follows that we can rewrite (\ref{bf1}) and
(\ref{bf2}) as
\begin{align}
\phi(x) &  =2^{n}\frac{\langle\phi|\psi\rangle}{\langle\psi|\gamma\rangle}%
\int\rho_{\phi,\psi}(z_{0})\widehat{T}_{\text{GR}}(z_{0})\gamma(x)\mathrm{d}%
^{2n}z_{0}\label{fi}\\
\psi(x) &  =2^{n}\frac{\langle\phi|\psi\rangle^{\ast}}{\langle\phi
|\gamma\rangle}\int\rho_{\phi,\psi}^{\ast}(z_{0})(z_{0})\widehat{T}%
_{\text{GR}}(z_{0})\gamma(x)\mathrm{d}^{2n}z_{0}.\label{psi}%
\end{align}

\section{Relation with Time-Frequency Analysis}

The Wigner formalism is widely used in time-frequency analysis (TFA), but
there is an interpretational and almost philosophical difference in the
viewpoints in QM and TFA. While the Wigner formalism is an essential tool for
expressing QM in its phase space version and leads to the Weyl quantization
scheme \cite{Birk,Birkbis,Littlejohn}, the situation is less clear-cut in TFA
where the appearance of cross-term correlations as $\mathcal{W}(\phi,\psi)$ is
an unwanted artefact; to eliminate or weaken these interference effects one
uses elements of the so-called Cohen \cite{Cohen,Gro} class (which would
typically be the Husimi transform in QM). Everything we have said above can be
re-expressed in terms of the cross-ambiguity function familiar from radar
theory (Woodward \cite{wood}, Binz and Pods \cite{Binz}), and related to the
Wigner transform by a symplectic Fourier transform:%
\[
\mathcal{A}(\phi,\psi)(z)=F_{\sigma}\mathcal{W}(\phi,\psi)(z)=F\mathcal{W}%
(\phi,\psi)(Jz)
\]
where $F$ is here the usual Fourier transform on $\mathbb{R}^{2n}$ and $J=%
\begin{pmatrix}
0 & I\\
-I & 0
\end{pmatrix}
$. It is given by the analytical expression%
\[
\mathcal{A}(\phi,\psi)(z)=\left(  \tfrac{1}{2\pi\hbar}\right)  ^{n}\int
e^{-\frac{\mathrm{i}}{\hbar}py}\phi^{\ast}(y+\tfrac{1}{2}x)\psi(y-\tfrac{1}%
{2}x)\mathrm{d}^{n}y.
\]
The symplectic Fourier transform being an involution we can rewrite the
definition (\ref{rofipsi}) of $\rho_{\phi,\psi}(x,p)$ in the form
\begin{equation}
\rho_{\phi,\psi}(x,p)=\frac{F_{\sigma}\mathcal{A}(\phi,\psi)(x,p)}{\langle
\phi|\psi\rangle}.\label{rofipsibis}%
\end{equation}
The use of this alternative approach does a priori not bring anything new from
a mathematical point of view, but could perhaps be useful for giving
interpretations of weak values and measurements in terms of the
\textquotedblleft source-target\textquotedblright\ formalism of radar theory.

\noindent\textbf{Acknowledgement 1}. The first author (MdG) has been funded by
the Austrian Research Agency FWF (Projektnummer P 23902-N13).

\noindent\textbf{Acknowledgement 2}. Both authors wish to express their thanks
to the Referees for a careful reading of our manuscript, and for having
pointed out several inaccuracies and shortcomings. Their comments have
substantially improved the final manuscript.

\end{document}